\documentclass[twocolumn,superscriptaddress,nobibnotes]{revtex4-1}

\usepackage{graphicx}
\usepackage{dcolumn}
\usepackage{bm}
\usepackage{times,mathptmx}
\usepackage{color}
\usepackage{multirow}
\usepackage{enumitem}
\usepackage{chngpage}
\usepackage{amsmath}

\begin{document}	
	
\title{\bf Integrated Pockels Laser}

\author{Mingxiao Li}
\thanks{These two authors contributed equally.}
\affiliation{Department of Electrical and Computer Engineering, University of Rochester, Rochester, NY 14627}
\author{Lin Chang}
\thanks{These two authors contributed equally.}
\affiliation{Department of Electrical and Computer Engineering, University of California Santa Barbara, Santa Barbara, CA 93106}
\author{Lue Wu}
\affiliation{T. J. Watson Laboratory of Applied Physics, California Institute of Technology, Pasadena, CA 91125}
\author{Jeremy Staffa}
\affiliation{Institute of Optics, University of Rochester, Rochester, NY 14627}	
\author{Jingwei Ling}
\affiliation{Department of Electrical and Computer Engineering, University of Rochester, Rochester, NY 14627}
\author{Usman A. Javid}
\affiliation{Institute of Optics, University of Rochester, Rochester, NY 14627}
\author{Yang He}
\affiliation{Department of Electrical and Computer Engineering, University of Rochester, Rochester, NY 14627}
\author{Raymond Lopez-rios}
\affiliation{Institute of Optics, University of Rochester, Rochester, NY 14627}
\author{Shixin Xue}
\affiliation{Department of Electrical and Computer Engineering, University of Rochester, Rochester, NY 14627}	
\author{Theodore J. Morin}
\affiliation{Department of Electrical and Computer Engineering, University of California Santa Barbara, Santa Barbara, CA 93106}
\author{Boqiang Shen}
\affiliation{T. J. Watson Laboratory of Applied Physics, California Institute of Technology, Pasadena, CA 91125}
\author{Heming Wang}
\affiliation{T. J. Watson Laboratory of Applied Physics, California Institute of Technology, Pasadena, CA 91125}
\author{Siwei Zeng}
\affiliation{Department of Electrical and Computer Engineering, Center for Optical Materials Science and Engineering Technologies, Clemson University, Clemson, SC 29634}
\author{Lin Zhu}
\affiliation{Department of Electrical and Computer Engineering, Center for Optical Materials Science and Engineering Technologies, Clemson University, Clemson, SC 29634}
\author{Kerry J. Vahala}
\email[]{vahala@caltech.edu}
\affiliation{T. J. Watson Laboratory of Applied Physics, California Institute of Technology, Pasadena, CA 91125}
\author{John E. Bowers}
\email[]{jbowers@ucsb.edu}
\affiliation{Department of Electrical and Computer Engineering, University of California Santa Barbara, Santa Barbara, CA 93106}
\author{Qiang Lin}
\email[]{qiang.lin@rochester.edu}
\affiliation{Department of Electrical and Computer Engineering, University of Rochester, Rochester, NY 14627}
\affiliation{Institute of Optics, University of Rochester, Rochester, NY 14627}


	
	
\begin{abstract}

	The development of integrated semiconductor lasers has miniaturized traditional bulky laser systems, enabling a wide range of photonic applications. A progression from pure III-V based lasers to III-V/external cavity structures has harnessed low-loss waveguides in different material systems, leading to significant improvements in laser coherence and stability. Despite these successes, however, key functions remain absent. In this work, we address a critical missing function by integrating the Pockels effect into a semiconductor laser. Using a hybrid integrated III-V/Lithium Niobate structure, we demonstrate several essential capabilities that have not existed in previous integrated lasers. These include a record-high frequency modulation speed of 2 exahertz/s (2.0$\times$10$^{18}$ Hz/s) and fast switching at 50 MHz, both of which are made possible by integration of the electro-optic effect.  Moreover, the device co-lases at infrared and visible frequencies via the second-harmonic frequency conversion process, the first such integrated multi-color laser. Combined with its narrow linewidth and wide tunability, this new type of integrated laser holds promise for many applications including LiDAR, microwave photonics, atomic physics, and AR/VR. 
	
	
\end{abstract}
	
\maketitle
\noindent\sffamily\textbf{Introduction}\nolinebreak
	
	\noindent\rmfamily The field of integrated semiconductor lasers has made many advances over the last few decades, spanning information technologies to fundamental science \cite{Spencer18,Xiang21,Newman19}. Using wafer-scale fabrication processes, these devices dramatically reduce the form factor of traditional bench-top laser equipment, and offer much lower power consumption and cost. Early laser designs were based entirely upon III-V semiconductors\cite{Hall62}, configured either as Fabry-Perot cavities emitting multiple wavelengths, or as distributed-feedback (DFB) designs for single frequency emission\cite{Yariv85, Coldren12}.  Besides providing coherent light generation across many applications, these devices serve as the key building block in on-chip systems by driving photonic integrated circuits (PICs) \cite{Smit11}. 
	
	With the continuing success of silicon photonics, integrated lasers have adopted passive cavities that are coupled to a III-V gain section. Endowed with enhanced photon lifetimes as well as reconfigurability, these integrated external-cavity-diode-laser (ECDL) structures \cite{Keyvaninia13, Hulme13, Tran19, Ishizaka09, Tanaka12}, by mimicking their bulk counterparts \cite{Fleming81, Liu81}, have significantly improved coherence and tunability in integrated photonics \cite{Tran19, Bergman18, Fan20,Jin21,Bowers19,Shen20, Nokia20, Sun20, Reed20, Cai21, Amir21}. Even more recently, with remarkable progress in fabrication of low loss Si/SiN waveguides \cite{Fan20,Jin21,Bowers19,Shen20}, the linewidths of integrated lasers are now comparable to those of state-of-the-art bench-top ECDLs and even fiber lasers. Such advances in coherence dramatically improves data capacity in communications \cite{Koizumi12} as well as accuracy in on-chip sensing and frequency metrology systems \cite{Spencer18}.

	\begin{figure*}[t!]
		\centering\includegraphics[width=2.0\columnwidth]{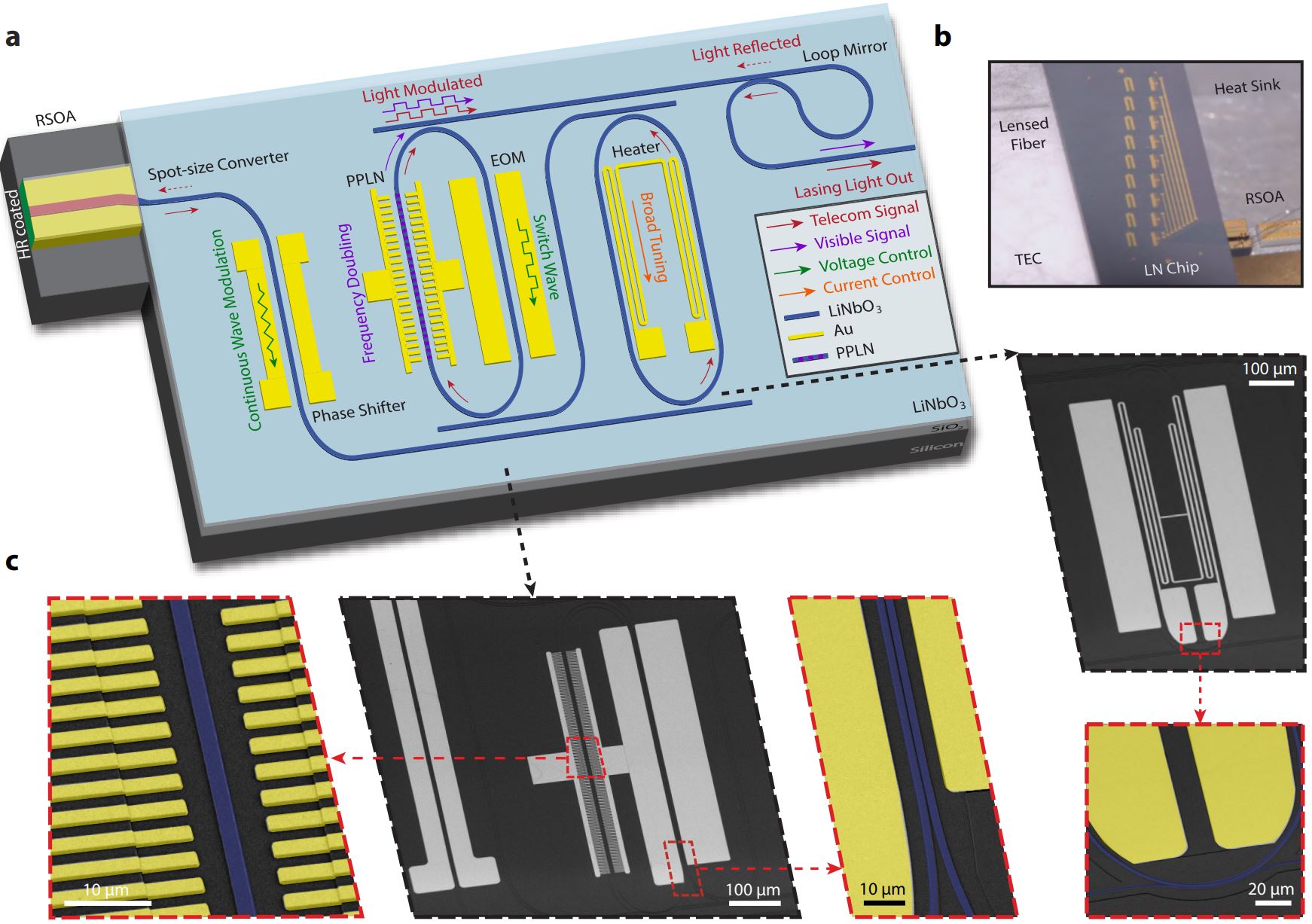}
        \caption{\label{Fig1} {\bf Concept and structure of the integrated Pockels laser.} {\bf (a)} Schematic of the hybrid integrated laser structure. {\bf (b)} Photo of the setup with an RSOA edge coupled to the device and sitting on heat sinks. A lensed fiber couples the light out from the device. The substrate is assembled by a thermoelectric controller for environmental temperature tuning. {\bf (c)}  Scanning electron microscope image of a fabricated device. False colors are applied to the zoomed-in figures highlighted by red dashed outlines.}
	\end{figure*}
	
		\begin{figure*}[t!]
		\centering\includegraphics[width=2.0\columnwidth]{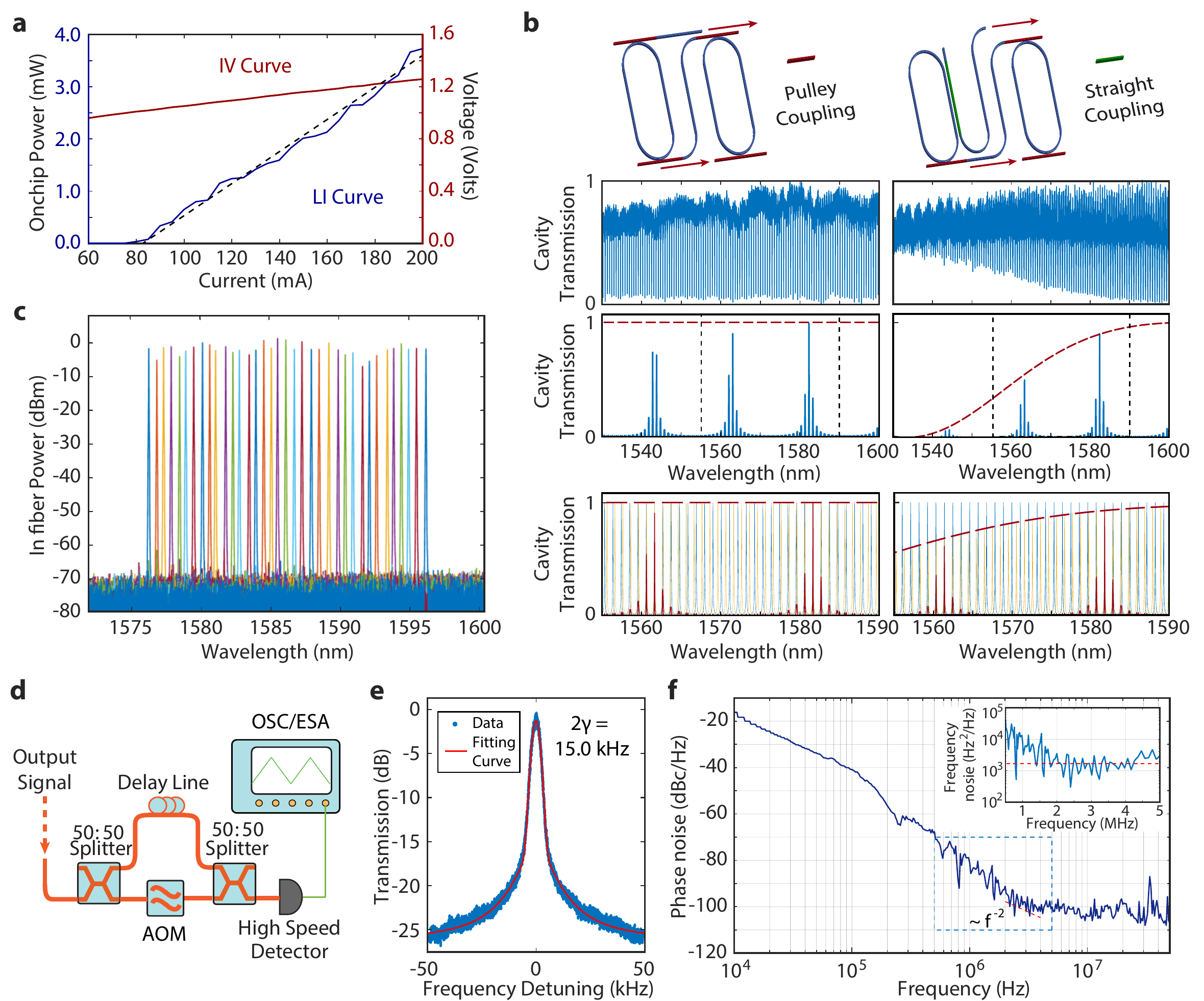}
		\caption{\label{Fig2} {\bf Power, linewidth, and wavelength tuning characteristics of the laser.} {\bf (a)} Measured LIV curve of the laser. {\bf (b)} Schematic of the Vernier-ring structure: the left one with pulley coupling and the right one with straight coupling replacing the output pulley bus waveguide. Second row: Measured transmission spectrum of both pulley and straight coupling.  Third row: Simulated transmission spectrum of the Vernier structure output. Left: with pulley coupling applied to all ports. Right: with output port replaced by straight coupling. Red dashed lines show the power transfer limited by the coupling rate. {\bf (c)} Thermal-optic tuning spectrum of the laser wavelength recorded by an optical spectrum analyzer.{\bf (d)} Setup schematic of delayed self-heterodyne phase noise measurement. AOM: acoustic-optical modulator. OSC: real-time oscilloscope. ESA: electrical signal analyzer. {\bf (e)} Noise spectrum of laser measured by sub-coherence delayed self-heterodyne measurement. The curve is fit by a combination of Lorentzian and Gaussian distribution. ${\gamma}$ is the half width of the half maximum of Lorentzian distribution. {\bf (f)} Noise spectra of laser measured by correlated delayed self-heterodyne measurement. The dark blue line is the phase noise signal measured by a real-time oscilloscope. The light blue line in the inset shows the frequency noise signal derived from the phase noise. A white noise floor is highlighted by the dashed red line correspondingly in both plots. The dashed rectangular box indicates the frequency range corresponding to the inset figure. }
	\end{figure*}
	
	However, despite these impressive achievements, key functions are missing in current integrated lasers. One outstanding problem lies in fast tuning and reconfigurability. Applications such as LiDAR require frequency modulation of a narrow-linewidth laser with high linearity and speed beyond the MHz level \cite{Riemensberger18}. In bench-top laser systems, these are usually realized by fast mechanical motion of components in an external cavity \cite{Fleming81, Liu81}, but similar strategies in integrated photonics are far more challenging. Most often, frequency tuning of integrated lasers relies on the thermo-optic (TO) effect \cite{Tran19}, which is relatively slow (kHz-level speed). And even while MHz-level frequency tuning can be achieved by current sweep of PN junctions of III-V or silicon waveguides, this carrier-induced effect produces unwanted intensity modulation \cite{Dhoore19} as well as additional loss that are not compatible with narrow linewidth lasers . This limitation is more severe at short wavelength below silicon's bandgap wavelength, where currently only thermal tuning can be used for external cavities of integrated lasers. As an example, in atomic physics, where switching speeds up to MHz level are required for ion/atom manipulation at visible and near visible bands, no integrated photonic solution of laser exist to provide fast reconfigurability. All current experiments have to rely on additional tuning provided by external modulators. In fact, even generating coherent light using integrated photonics in these wavelength is quite challenging. In contrast to free-space laser cavities where nonlinear media can be readily implemented within the resonator to generate short-wavelength light by frequency conversion \cite{KOZLOVSKY88, Xue06}, an integrated nonlinear cavity suitable for electrical pumping has so far remained elusive. In these visible/near-visible applications, integrated photonics must rely on very challenging design, growth, and processing developments using new gain media. The resulting difficulties have presented a bottleneck to on-chip solutions in a wide range of evolving applications.  
	
	In this work, we propose and demonstrate a new family of lasers, the Pockels laser, to fill these long-standing gaps in the integrated photonics paradigm. By using lithium-niobate-on-insulator (LNOI) waveguide elements to form an external cavity,  we unite a III-V gain section with the Pockels effect in an integrated laser. This adds several new capabilities to the toolbox of on-chip lasers including fast on-chip reconfigurability of a narrow linewdith laser (fundamental lineweidth ~ 11.3 kHz) with laser-frequency tuning at a record speed of 2.0 EHz/s, as well as fast switching at a rate of 50 MHz. On account of the low required drive voltages, these functions can be directly driven by CMOS signals.  Furthermore, using an intracavity periodically-poled lithium niobate (PPLN) waveguide section embedded in one of the Vernier rings, we demonstrate the first multi-color integrated laser that emits high-coherence light at telecommunication wavelengths and in the visible band. It is also the first narrow-linewidth laser with fast reconfigurability at visible.

\noindent\sffamily\textbf{Results}\nolinebreak
	
\noindent\rmfamily\textbf{Laser design.} The laser structure is shown in Fig.~\ref{Fig1}a, where a III-V reflective semiconductor optical amplifier (RSOA) is edge-coupled to an external cavity on a LNOI chip, forming a hybrid integrated laser \cite{Ishizaka09, Tanaka12}. Lithium niobate (LN) is well-known for its superior capability in optical modulation and frequency conversion \cite{Fejer95, Bossi00, Arnan18, Loncar21}. A laser cavity built upon would enable intriguing laser functionalities significantly beyond the reach of conventional integrated lasers, as we will show in detail below. To avoid mode mismatch between the RSOA and LNOI chip, a spot-size converter is adopted in the system, obtaining a minimal mode mismatch between a III-V waveguide and a 5-${\mu}$m-wide 600-nm-thick LNOI waveguide, whose mode profiles are shown in the Methods. To minimize facet reflection, the III-V facet is coated with anti-reflective (AR) layers, and LNOI's input-facet coupling waveguide is angled by 10 degrees to achieve a reduced reflectivity (around 10\% simulated by FDTD Lumerical) and match the angle of injected light. The reflectivity can be further reduced by applying AR coating to LNOI's input-facet.

The LNOI external cavity is a Vernier mirror structure consisting of two racetrack resonators. The geometry of racetracks and bus waveguides are tailored to minimize the number of coupled mode families to avoid multi-mode lasing. The coupling is carefully selected by taking the lasing power, laser linewidth, and the tuning speed of the cavity into consideration. The free spectral range (FSR) of the resonators is set to be 70 GHz, with a 2 GHz difference between the two resonators, which corresponds to a Vernier FSR of 2.4 THz \cite{Tran19}. The shape of the racetrack is optimized with the trade-off between the EO modulation efficiency that requires a long straight section, and the optical scattering loss that requires a large curvature radius. As a result, an Euler curve profile is employed to minimize the scattering loss while maximizing the length of the racetrack. The polarization of the fundamental quasi-transverse-electric (quasi-TE) mode is aligned to harness the large Pockels effect of LN ($r_{33}$ = 30 pm/V, $d_{33}$ = 19.5 pm/V \cite{Nikogosyan05}) at the straight section of the racetrack. 
	
To combine versatile functions into one laser structure, each of the resonators is designed for a different purpose. The first one is incorporated with a micro heater for broad wavelength tuning via the TO effect, while the second one is integrated with driving electrodes designed for high-speed EO tuning. Moreover, the second resonator is tailored to be compatible with the SHG process, with a PPLN section embedded directly inside the resonator. Furthermore, a tunable phase control section is also implemented in the cavity to align the longitudinal laser cavity mode with the Vernier mode \cite{Tran19}. Benefiting from the strong EO Pockels effect in LN, the phase-control section is operated via the EO effect instead of the commonly used TO effect \cite{Tran19}. In contrast to the conventional TO approach that is slow (kHz-speed), power hungry and suffers from the thermal crosstalk problem, the EO Pockels approach enables high-speed, energy efficient, and independent control of individual functionalities as we will show below. Finally, a Sagnac loop ring is placed at the end of the device to function as the output end mirror of the laser cavity. The output-facet waveguide is designed for optimized coupling to a tapered fiber for performance characterization, as shown in Fig.~\ref{Fig1}b.

\medskip
	
\noindent\textbf{Linear performance.} A fabricated device is shown in Fig.~\ref{Fig1}c (see Methods for the details of device fabrication). The intrinsic quality (Q) factor of the racetrack resonators is around 1.2 million, while the external coupling Q is much lower, varying from 5.0$\times$10$^{4}$ to 1.5$\times$10$^{5}$, which determines the loaded Q of the device. The laser light-current-voltage (LIV) curve measurement is performed for the lasing mode at 1581.12 nm, which has a threshold current of 80 mA and an on-chip power of around 3.7 mW at 200 mA, 
as shown in Fig.~\ref{Fig2}a. The highest on-chip power measured from this device can reach more than 5.5 mW by adjusting the Vernier mirror conditions.

	\begin{figure*}[t!]
		\includegraphics[width=2.0\columnwidth]{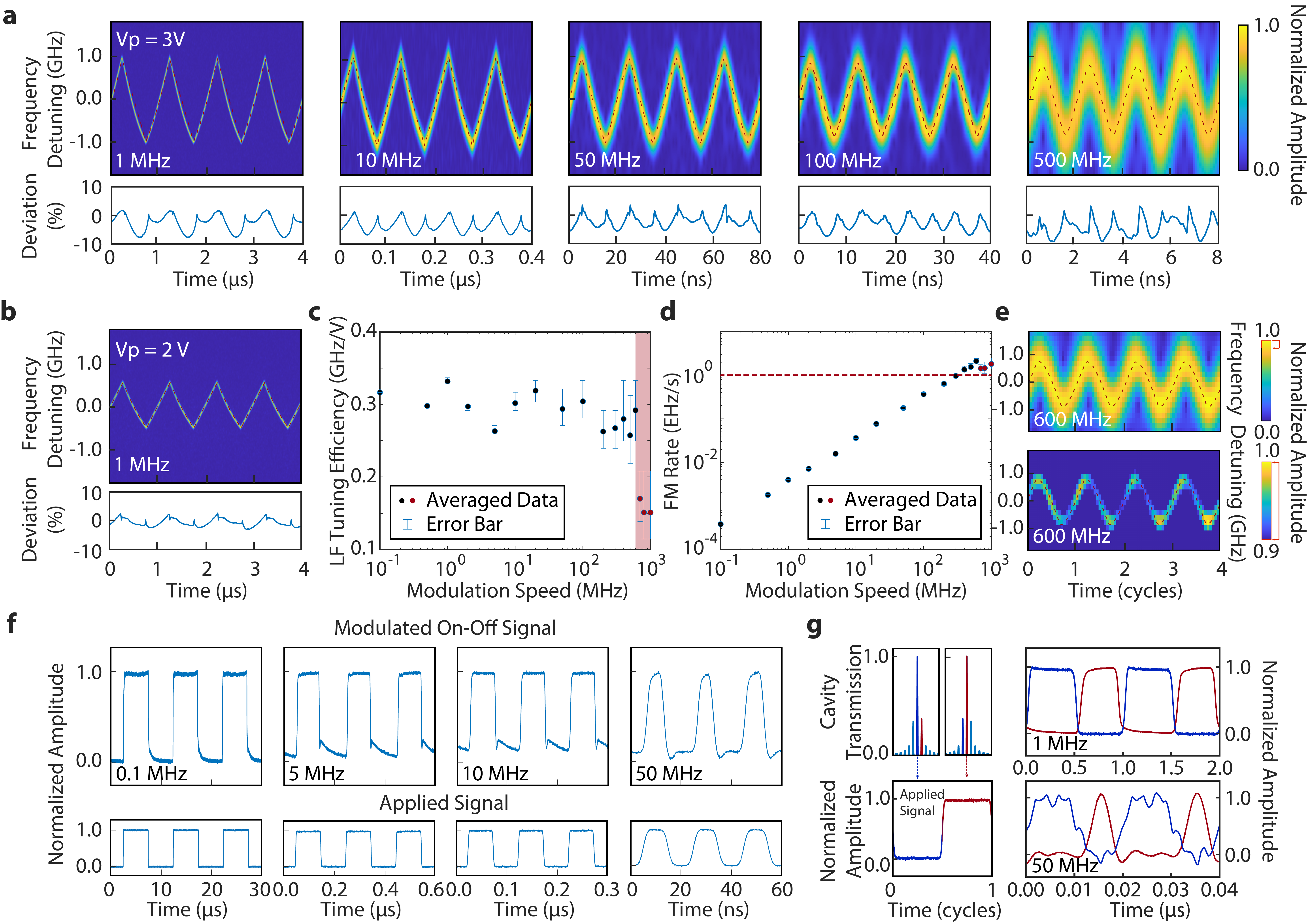}
		\caption{\label{Fig3}  {\bf High-speed tuning and switching characteristics of the laser.} {\bf (a)} Time-frequency spectrograms of the beat note between the Pockels laser and a reference diode laser, at different modulation frequencies. The red dashed lines show the triangular waveforms of the driving electrical signal (with an amplitude of $V_P$ = 3~V ) (The slight waveform distortion is induced by the limited bandwidth of a RF amplifier used to boost the electrical signal. See Methods for details). The lower panel shows the deviation of signal compared to the modulating triangular ramp. {\bf (b)} Time-frequency spectrogram of the beat note with a smaller driving voltage of $V_P$ = 2 V. {\bf (c)} The recorded laser-frequency (LF) tuning efficiency of the laser frequency. The red shaded area indicates the frequency range beyond the photon lifetime limit of the laser cavity. {\bf (d)} Recorded laser frequency modulation rate. Red dashed line highlights the level of 1 EHz/s. {\bf (e)} Time-frequency spectrogram of the beat note signal with a modulation frequency of 600~MHz and a driving voltage of $V_P$ = 3~V. The lower one shows the same spectrogram but with a zoom-in amplitude range of 90--100\%. {\bf (f)} On-off intensity switching waveform of the laser (top row) at different modulation frequencies. The corresponding driving electrical signal is shown on the bottom row, with an amplitude of $V_P$ = 3~V.  {\bf (f)} Left: Schematic shows the switching between adjacent Vernier lasing mode in red and dark blue curve with the applied electrical signal shown on the bottom. Right: Recorded waveforms of the two lasing modes, at a modulation frequency of 1~MHz and 50~MHz.}
	\end{figure*}
	
	The two racetrack resonators use different coupling structures: the first one uses a Pulley coupler, but the second racetrack adopts a straight waveguide coupler at both telecom and near-infrared wavelengths for SHG operation. Both bus waveguides are designed to work only for the fundamental quasi-TE mode. Fig. \ref{Fig2}b shows the details of these two structures, together with their recorded transmission spectra. A pulley coupling structure benefits the bandwidth of the lasing spectrum, but also raises the risk of multi-mode lasing (Fig.~\ref{Fig2}b, lower left). Here, the use of the straight coupling design for one resonator significantly suppresses the mode that is one Vernier FSR away  (Fig.~\ref{Fig2}b, lower right). With this design, single-mode laser is achieved, with a high side-mode suppression ratio (SMSR) greater than 50~dB, as shown in Fig.~\ref{Fig2}c. The coarse wavelength tuning is realized by thermo-optical tuning a Vernier ring resonator as described above, with a tuning range of $\sim$20 nm from 1576 nm to 1596 nm which agrees with the designed Vernier FSR (2.4 THz). The high SMSR is maintained over the entire tuning range.

	The linewidth of the laser is first characterized by a delayed self-heterodyne method \cite{Linewidth86}, with a setup shown in Fig.~\ref{Fig2}d (details in Methods). The recorded data is fit by a combination of Lorentzian and Gaussian distributions, resulting in a Lorentzian linewidth of 15.0~kHz, as shown in Fig.~\ref{Fig2}e. To confirm the linewidth, we further applied the correlated delayed self-heterodyne phase noise method \cite{Jin21,Linewidth08} (details in Methods). The recorded phase noise is shown in Fig.~\ref{Fig2}f, with the corresponding frequency noise shown in the inset. The white noise floor of $\sim$1.8$\times$10$^{3}$ ${\rm Hz^2/Hz}$ is found at a frequency around 3 MHz, which corresponds to an intrinsic linewidth (or Lorentzian linewidth) of 11.3~kHz for the laser, further confirming the narrow-linewidth performance.

\medskip

\begin{figure*}[t!]
    \includegraphics[width=2.0\columnwidth]{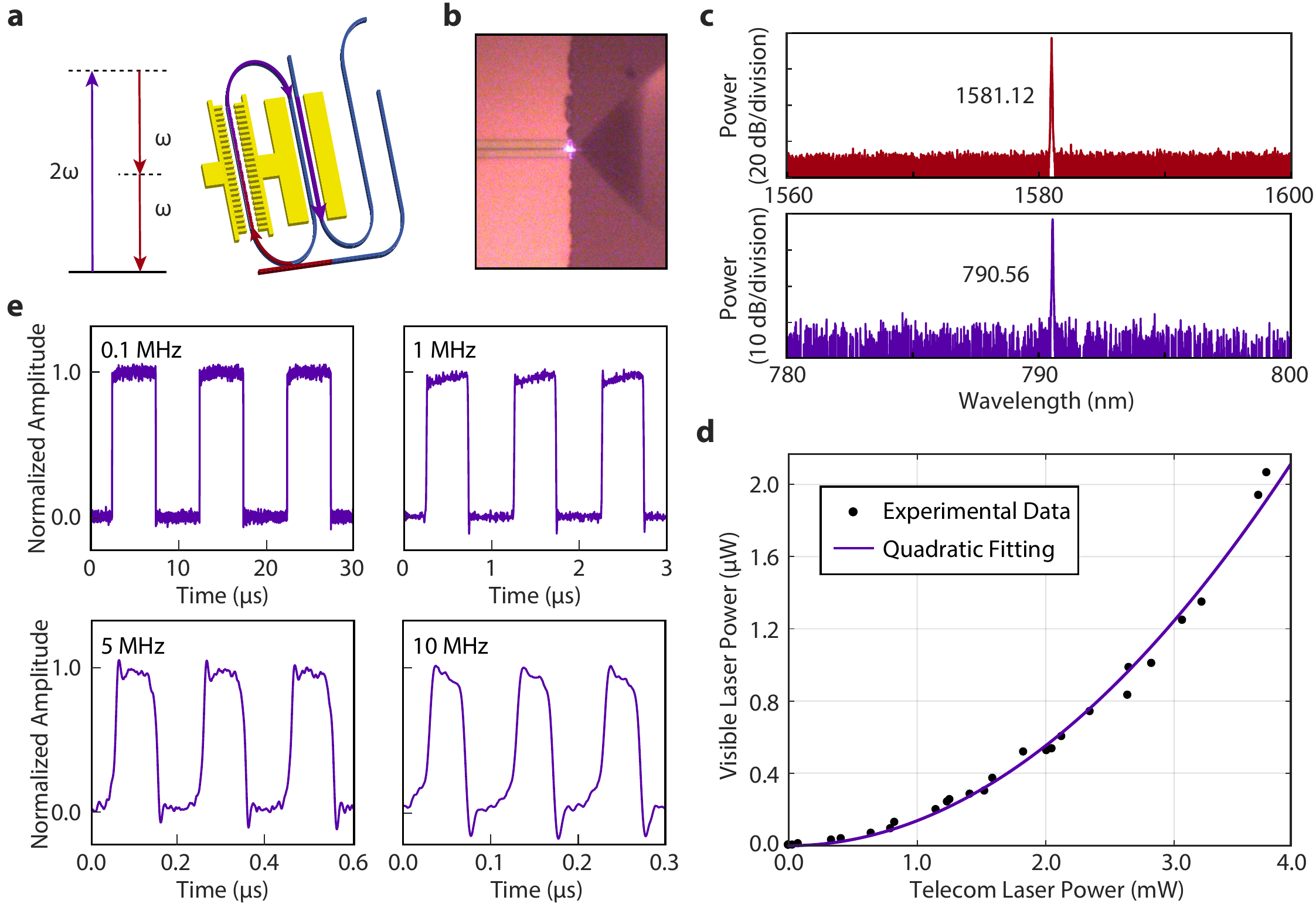}
	\caption{\label{Fig4} {\bf Performance of the dual-wavelength laser.} {\bf (a)} Schematic of the SHG process implemented in the resonator. {\bf (b)} Optical microscope image of the produced SHG light at the output facet of the laser chip. {\bf (c)} Optical spectra of the fundamental-frequency lasing mode (top) and the up-converted light at the second harmonic (bottom). {\bf (d)} Recorded power dependence between the laser outputs at the telecom wavelength and in the visible. The solid curve shows a quadratic fitting to the recorded data (solid dots). {\bf (e)} On-off switching waveform of the SH light at different modulation frequencies. }
\end{figure*}

\noindent\textbf{Ultrafast frequency tuning and switching of the laser.} There is an urgent demand for high-speed frequency modulation (chirping) of a narrow linewidth laser in various applications. For example, the current FMCW LiDAR uses MHz-level laser-frequency-modulation rate with GHz-level chirping range. Even higher speed reconfigurability is demanded in applications such as the frequency modulation spectroscopy \cite{Bjorklund1980}. The Pockels laser is ideal for this purpose, where the laser frequency can be fast tuned by electro-optically tuning the phase shifter section. As such EO tuning of the phase shifter only changes the effective optical path length of the laser cavity while without introducing any loss, it offers an elegant approach for fast frequency tuning without linewidth degradation or parasitic intensity modulation. 

To show this capability, we apply a high-speed driving electrical signal to the phase shifter and monitor the laser-frequency tuning by beating the laser ouput with a reference narrow-linewidth laser whose frequency has a 8.0-GHz initial offset (details in Methods). For a better illustration of the fine tuning performance via the Pockels effect, the driving electrical signal is in a triangular waveform with a modulation frequency ranging from 0.1 MHz to 1 GHz and an amplitude of $V_P$ = 3.0~V. The heterodyne beat note is recorded and processed by a short time Fourier transform (STFT). Recorded data are shown in Fig.~\ref{Fig3}a for the modulation frequency ranging from 1 MHz to 500 MHz, together with the deviation of laser-frequency modulation linearity (details in Methods). As shown in the figure, the waveform of laser-frequency modulation follows faithfully that of the driving electrical signal (dashed curves) at all frequencies, with a nonlinearity less than 10\% for the modulation frequency up to 500 MHz. As shown in Fig.~\ref{Fig3}b, with a lower driving voltage of $V_P$ = 2.0 V, the nonlinearity of laser-frequency modulation can be reduced to 3\%. In Fig.~\ref{Fig3}a, the resolution of the spectrogram degrades with increased modulation speed, which is simply due to the limited sampling rate of the oscilloscope used to record the laser beating signal. 

As shown in Fig.~\ref{Fig3}c, the amplitude of laser-frequency modulation remains at a fairly constant level across the broad range of modulation frequency up to 600~MHz, with a value in the range of (1.6 -- 2.0)~GHz that corresponds to a tuning efficiency of (0.26 -- 0.34)~GHz/V. As a result, the laser-frequency modulation rate increases nearly linearly with modulation speed, as shown in Fig.~\ref{Fig3}d, reaching a value of 2.0~EHz/s (2.0$\times$10$^{18}$ Hz/s) at the modulation frequency of 600 MHz. The frequency modulation rate starts to saturate when the modulation frequency increases beyond 700 MHz, simply because the modulation speed reaches the photon lifetime limit of the laser cavity (estimated to be $\sim$0.2~ns), leading to a degradation of EO tuning efficiency (Fig.~\ref{Fig3}c).

The laser-frequency modulation of our laser is fairly independent of intensity modulation, since with direct phase modulation inside the cavity, the intensity variation of the laser is merely caused by the mode mismatch between the cavity longitudinal mode and the Vernier mode, which is fairly small within the bandwidth of the Vernier mode. This feature is shown in Fig.~\ref{Fig3}e, which also plots the time-frequency spectrogram with a zoom-in amplitude range. It shows clearly that a continuous signal is observed, with a small amplitude variation less than 10 \%. This is in strong contrast to other frequency modulation approaches such as current modulation of diode laser that suffer from the considerable coexisting intensity modulation, undermining the quality of related applications. The residual intensity modulation in our laser can be further suppressed by a coordinated EO tuning of both the phase shifter and the Vernier ring resonator.

In addition to the pure frequency modulation shown above, the Pockels laser also allows a fast on-off switching of the lasing mode. This pure intensity modulation is realized by applying a square wave to electro-optically drive the racetrack resonator (instead of the phase shifter as done above. Details in Methods). The consequential mode mismatch between two resonators introduces rapid degradation of Vernier mode, resulting in drastic change of the intra-cavity loss, which enables an on-off behavior of laser, acting as a high-speed switch. As shown in Fig.~\ref{Fig3}f, with applied modulating frequencies ranging from 0.1 MHz to 50 MHz, both on- and off-states can be observed distinctly with a 10 \% -- 90 \% rise and fall times around 3~ns, limited by the speed of the applied driving signal (see Methods). The switching performance degrades when the modulation frequency increases beyond 50 MHz, which is likely due to the oscillatory nature of the laser during cavity mode stabilization \cite{Nokia20}. 

Further increase of the amplitude of the driving electrical signal would trigger the second adjacent lasing mode, leading to intriguing fast laser mode switching. Fig. \ref{Fig3}g shows this phenomenon, where the amplitude of the driving square wave is increased to $V_P$ = 4~V, a value adequate to switch the laser between two lasing modes with one FSR apart. To observe the switching behavior, the laser output is separated by a wavelength-division demultiplexer (WDM) into two channels at different wavelengths to monitor the dynamics of the individual lasing modes (details in Methods). As shown in Fig.~\ref{Fig3}g, the switching between the adjacent laser modes is observed with a clear rising edge around 3 ns. The quality of signal is limited by the requirements of synchronous control on the phase shifter for longitudinal mode alignment, which can be implemented in future work. The fast wavelength switching demonstrated here is of great potential for application in data communication and access network. 
	
	\medskip
	
	\noindent\textbf{Dual wavelength laser.} In traditional integrated photonics, SHG can only be pumped using an external laser, which is complicated in operation, difficult to achieve fast reconfigurability.  Here, for the first time, we incorporate PPLN directly into the integrated laser cavity, which enables inherent SHG by the integrated laser itself, significantly reducing the system complexity as shown in Fig.~\ref{Fig4}a. Moreover, the strong intracavity laser power compared to the laser output can further enhance the SHG process. The resonance matching between the fundamental frequency (FF) and second harmonic (SH) modes for the SHG process is precisely controlled by the temperature of the laser chip (TEC in Fig.~\ref{Fig1}b). As soon as the device starts to lase at 1581.12 nm, the produced SH is readily visible at the output facet of the laser chip, as shown in Fig.~\ref{Fig4}b. The spectra of the fundamental telecom laser and the frequency-doubled visible wave are shown in Fig.~\ref{Fig4}c, showing a dual-wavelength lasing behavior. The recorded laser output at both wavelengths are plotted in Fig.~\ref{Fig4}d, which shows a clear quadratic power dependence between the two colors, an intrinsic nature of the SHG process.

	One great advantage of Pockels laser is the capability to incoporate wavelength converters inside the laser cavity, thus offers fast reconfigurability of the visible light simply by manipulating its telecom pump laser as shown in the previous section. Here we show the MHz-level switching of visible light that is particularly important for atomic/ion trapping experiments to conduct the imaging light controlling, optical pumping and brief laser cooling steps, but didn't exist in previous integrated photonic approaches. In experiment, we apply a square-wave driving signal to modulate the lasing cavity, as done in Fig.~\ref{Fig3}f, and monitor the waveform of the frequency-doubled light (details in Methods). As shown in Fig.~\ref{Fig4}e, the on-off switching is clearly observed with a modulation frequency from 0.1~MHz to 10~MHz, with waveforms closely following those shown in Fig.~\ref{Fig3}f. Such a switching speed can satisfy the speed requirement of almost all the atom/ion manipulating experiment.

    The current two-color laser exhibits a wavelength tuability of $\sim$ 10~nm. The details are provided in the Method section. The intracavity PPLN can be further engineered for broadband phase matching to cover the entire lasing wavelength range of the fundamental wave, which would enable a broadly tunable operation of the two-color laser. On the other hand, the power of the produced SH light can be further increased by optimizing the coupling efficiency between the RSOA and the LN chip, which is currently relatively low and limits the intracavity lasing power at the fundamental frequency. The SH power can also be increased by implementing longer sections of PPLN inside the cavity to enhance the SHG efficiency. Moreover, the SHG efficiency can be increased by optimizing the external coupling Q of the resonator. For example, a 10-fold increase in external coupling Q to 0.6 million will lead to an SHG conversion efficiency of 15 \% that corresponds to more than 1 mW SH power.

\bigskip

	\noindent\sffamily\textbf{Discussion}\nolinebreak

	\noindent\rmfamily Besides the performances we presented here, the implementation of the Pockels effect into integrated laser can lead to more novel functionalities compared to previous integrated lasers. The capability of fast laser-frequency reconfigurablitiy by the EO effect, combined with the intensity modulation by varying the current, potentially can enable fully integrated optical arbitrary waveform generator (AWG) on chip for communications and microwave photonics. The cavity design can be further optimized by engineering the quality factors of the ring resonators to support much higher speed modulation, while maintaining a narrow linewidth at the same level with those of current ECDLs. Furthermore, by changing the design of the PPLN inside the resonator, the pump can be frequency converted to a much broader spectrum range, through cascaded sum frequency generation to shorter wavelength at green or blue, or optical parametric oscillator to mid-IR wavelength. Such flexible wavelength generation on-chip can significantly relieve the difficulties in material growth and device processing of different laser epi structures. We also expect that, with the advance of fabrication in increasing the integration level (heterogeneous integration), a fully integrated, fourndry-based solution for this new type of laser will show up in the near future.
	
	In summary, by hybrid integration of a LN external cavity with a III-V RSOA, we demonstrated the first integrated Pockels laser. The device exhibits a great reconfigurability based on the EO effect, featuring a record-high laser-frequency modulation speed of 2.0 EHz/s and switching speed up to 50 MHz. This exceptional performance affords a promising solution to LiDAR and many other applications. Moreover, by incorporating the high nonlinear frequency conversion capability of LN, the first integrated multi-color laser with telecom and its SHG wavelength output is realized. The further combination of these two functions helps to demonstrate fast switching of the wavelength converter with up to 10 MHz speed, paving the path to applications of integrated light sources for atomic physics, AR/VR and sensing.
	
    The demonstrations in this work not only extend the applications of the LNOI platform, but more generally, provide a solution to various problems in nanophotonics. They also provide a design path to multi-color fully integrated systems with various functionalities. Such systems have many potential applications in nonlinear optics, optical signal processing systems, quantum photonics and optical communications.
	
    \bigskip
    
\begin{figure*}[t!]
	\includegraphics[width=2.0\columnwidth]{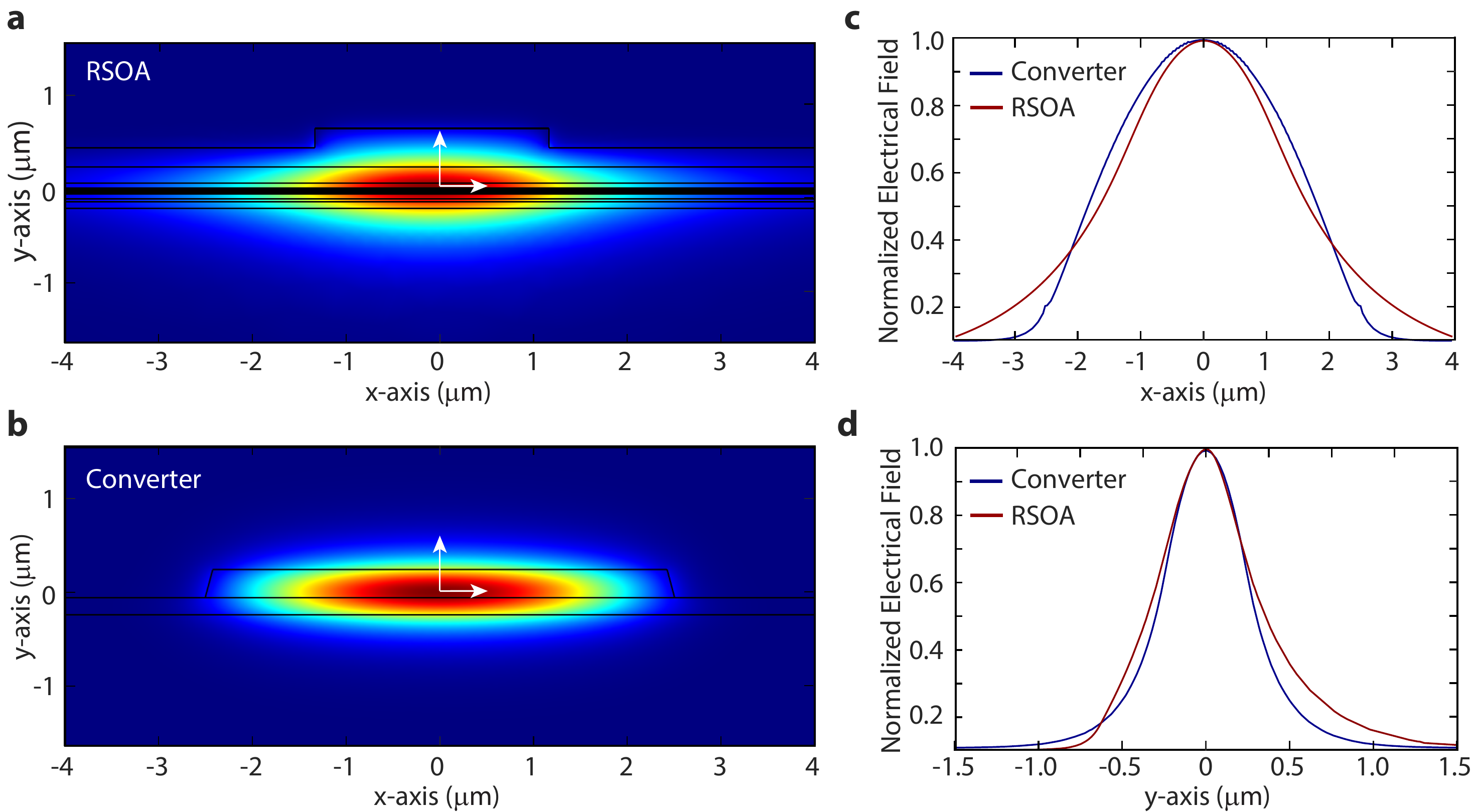}
	\caption{\label{Extended_Fig1}  {\bf Design of the spot-size converter.} {\bf (a, b)} The simulated normalized electrical field of the mode profile from a typical RSOA gain chip and the designed converter respectively. {\bf (c, d)} The normalized electrical field of the simulated mode profile along x- and y-directions at the center of modes as labeled with white arrows in {\bf (a, b)} respectively.}
\end{figure*}

\begin{figure*}[t!]
	\includegraphics[width=2.0\columnwidth]{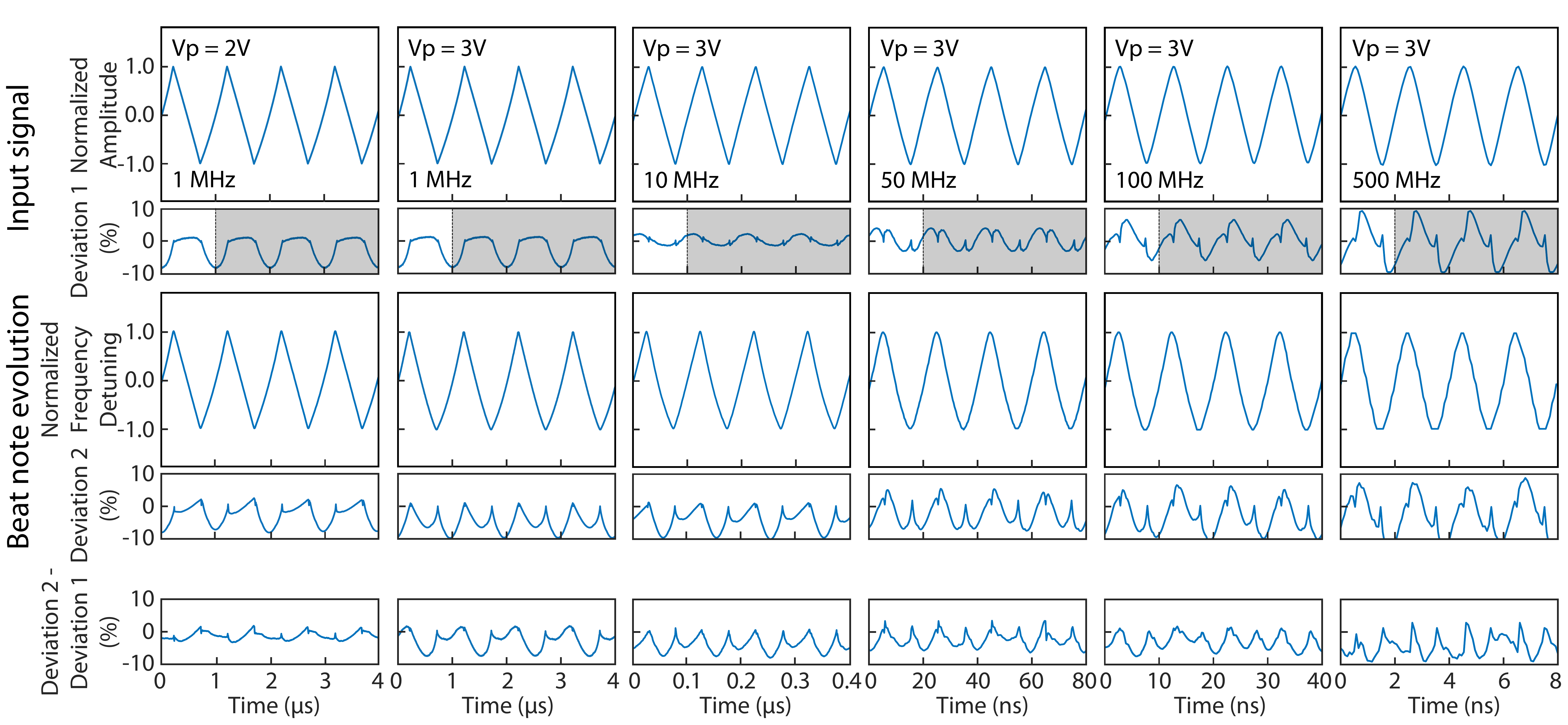}
	\caption{\label{Extended_Fig2} {\bf Characterization of the linearity of laser frequency modulation.} The first row shows the recorded waveform of the electrical signal used to drive the phase shifter, together with its deviation to a perfect triangular waveform (induced by an RF amplifier) defined as \emph{Deviation 1}. Because of the periodicity of the signal, \emph{Deviation 1} is fully represented within one modulation period, which is obtained by averaging it over multiple modulation periods to have a better accuracy. The periodic waveform of \emph{Deviation 1} is thus obtained by duplicating it over multiple modulation periods in time, as shown in the shaded region. The second row shows the waveform of laser frequency modulation retrieved with STFT from the recorded laser beat note. \emph{Deviation 2} shows its difference from a perfect triangular function, normalized by the peak-peak amplitude. The last row plots the difference between \emph{Deviation 2} and \emph{Deviation 1}. }
\end{figure*}

\begin{figure*}[t!]
	\includegraphics[width=2.0\columnwidth]{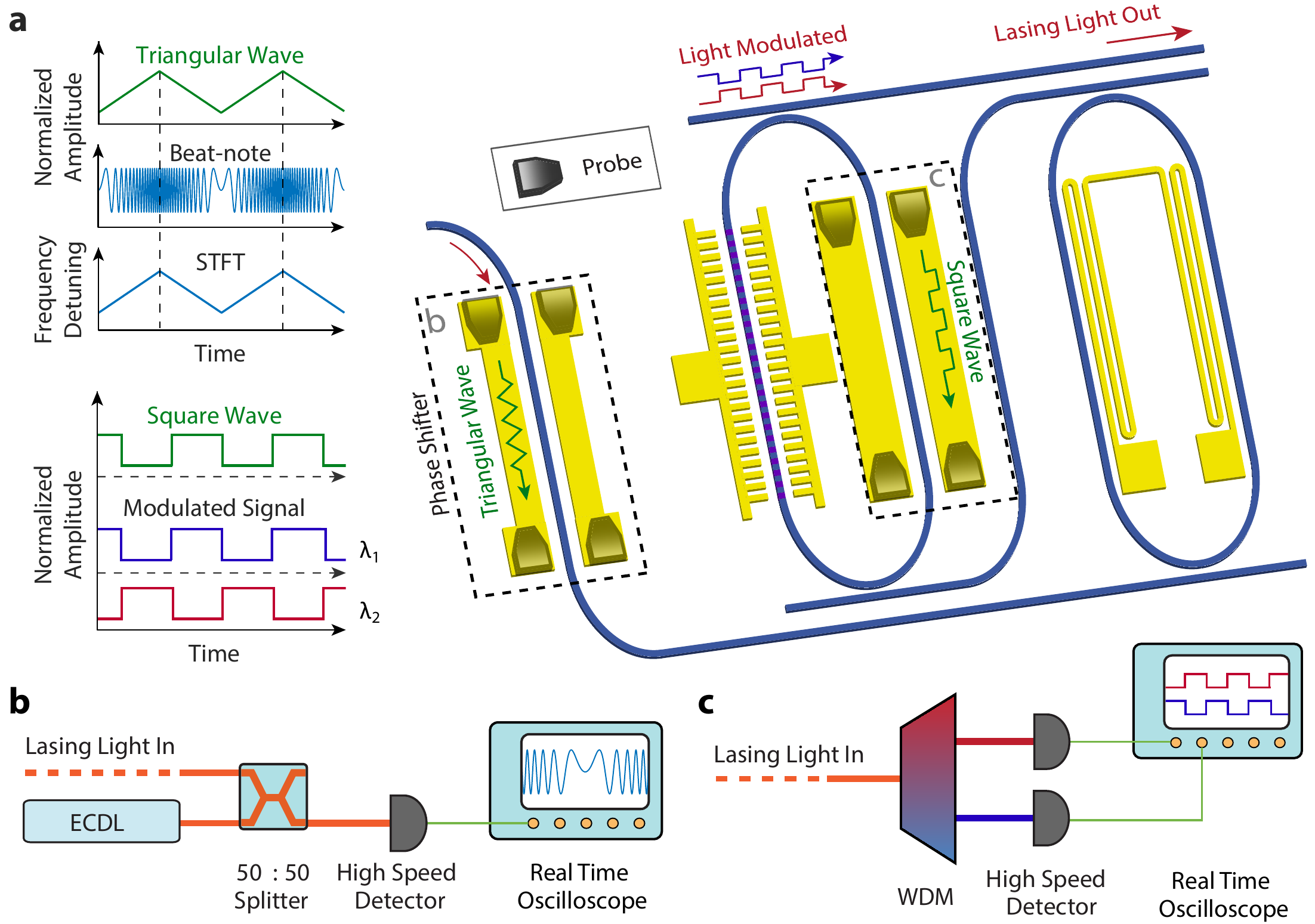}
	\caption{\label{Extended_Fig3} {\bf Experimental setup for high-speed tuning and switching of the laser.} {\bf (a)} Right: Schematic of device with probes placed for high-speed characterization. Dashed lines are used to highlight the operating units. Up Left: Schematic of signal waveforms for laser frequency modulation. Top: triangular electrical signal to drive the phase shifter; middle: laser beat note; bottom:  time-frequency spectrogram retrieved by the STFT. Down Left: Schematic of signal waveforms for laser-mode switching. Top: square-wave electrical signal to drive the Vernier ring resonator; Bottom: the produced waveforms at two lasing modes. STFT: short time Fourier transform. {\bf (b,c)} Schematic of experimental setup to record the laser beat note and the laser-mode switching, respectively. ECDL: external cavity diode laser; WDM: Wavelength-division multiplexer.}
\end{figure*}

    \noindent\sffamily\textbf{Methods}\nolinebreak
    
    \noindent\rmfamily\textbf{Device fabrication.} The devices were fabricated on a $600$-nm-thick x-cut single-crystalline LN thin film bonded on a 4.7-${\mu}$m silicon dioxide layer sitting on a silicon substrate (from NanoLN). The waveguide and racetrack structures are patterned by ZEP-$520$A positive resist via electron-beam lithography; an Ar$^+$ plasma milling process is used to  transfer the pattern to the LN layer with the etch depth of 300-nm. The resist was removed by the solvent 1165 resist remover afterward. The metal electrode layer (10 nm Ti/500 nm Au) was patterned by PMMA and deposited by an electron-beam evaporator, then formed by an overnight lift-off process. Finally, the devices were diced and polished to minimize the edge coupling loss.
    
    \medskip

	\noindent\rmfamily\textbf{Spot size converter.} The mode mismatch can cause enormous insertion loss between the RSOA and LNOI chip, which degrades the output power of laser seriously. To resolve this potential issue, we introduce a spot-size converter to the system to match the mode profile at the edge of the RSOA. A minimal mode mismatch is found by the implementation of an input waveguide with 5 ${\mu}$m width and 600 nm thickness of LN thin film, simulated by an FEM software, as shown in extended Fig.~\ref{Extended_Fig1}. With an etch depth of LN over 200 nm, minor variation to the matching efficiency is observed, allowing us to employ it to various designs simply.   
	
	\medskip

\noindent\rmfamily\textbf{Laser linewidth measurement.} As shown in Fig.~\ref{Fig2}d, the lasing signal passes a 50:50 splitter. One optical path is delayed by a 40-km-long fiber delay line, while the other is modulated by an acoustic optic modulator (AOM) at 55 MHz. The 40 km delay line realizes a low linewidth measurement down to 5 kHz, while the AOM is used to apply a frequency detuning to the signal, resulting in a radio frequency after recombination, which is relatively free from technical noise from electronics, vibrations, and other environmental factors. The signals are recombined by a 50:50 coupler and detected by a high-speed photodetector, then analyzed by an electrical signal analyzer. The recorded data is fit by a combination of Lorentzian and Gaussian distributions, resulting in a Lorentzian linewidth of 15.0~kHz, as shown in Fig.~\ref{Fig2}e. 
	
For the correlated delayed self-heterodyne phase noise measurement \cite{Jin21,Linewidth08}. The experiment setup is similar to the previous one as shown in Fig.~\ref{Fig2}d, except that the optical path is delayed by a 1-km-long fiber line, and the signal is processed using a real-time oscilloscope. The measured phase noise is shown in Fig.~\ref{Fig2}f. The frequency noise is calculated from the phase noise, as shown in the inset of Fig.~\ref{Fig2}f. The white noise floor of $\sim$1.8$\times$10$^{3}$ ${\rm Hz^2/Hz}$ is found at a frequency around 5 MHz. This value can be multiplied by 2${\pi}$ to indicate the intrinsic linewidth, or Lorentzian linewidth of the laser, which is 11.3 kHz in this case. The loaded optical Q of this device is measured to be $\sim$7.0$\times$10$^{4}$. As a comparison, we recorded a larger laser linewidth of 41~kHz for another laser device with a lower optical Q of $\sim$2.5$\times$10$^{4}$. A higher loaded Q results in a narrower laser linewidth, which can be further improved in future work.

\medskip

\noindent\rmfamily\textbf{Characterization of the linearity of laser frequency modulation.} The laser frequency modulation is realized by electro-optically modulating the phase shifter of the device. The electrical driving signal with a triangular waveform is generated by a high-speed arbitrary waveform generator (AWG) (Keysight M8196A) and is amplified to aimed voltage amplitudes by a wide-band RF amplifier before it is applied to the phase shifter. Its waveform is shown in the first row of Fig.~\ref{Extended_Fig3}. Due to the limited bandwidth, the RF amplifier introduces certain distortions to the signal waveform. We first characterize such distortion by comparing the signal waveform with a perfect triangular waveform. Their difference, normalized by the peak-peak amplitude, is defined as \emph{Deviation 1} and is shown in the first row of Fig.~\ref{Extended_Fig3}, which quantifies the magnitude of waveform distortion of the electrical driving signal. \emph{Deviation 1} serves as the reference to characterize the linearity of laser frequency modulation (see below). Since the signal waveform is purely periodic, the waveform of \emph{Deviation 1} can be fully represented within one modulation period. To have a better accuracy, it is obtained by averaging over multiple modulation periods. 


The beat note between the Pockels laser and the reference ECDL is recorded by a real-time oscilloscope (Keysight UXR0334A), as illustrated in Fig.~\ref{Extended_Fig2}b. The recorded signal is processed by STFT to retrieve the time-frequency spectrogram which is shown in the second row of extended Fig.~\ref{Extended_Fig3}. It is compared with a perfect triangular waveform, and their difference, normalized by the peak-peak amplitude, gives \emph{Deviation 2} which is also shown in the second row of Fig.~\ref{Extended_Fig3}. The difference between \emph{Deviation 2} and \emph{Deviation 1} thus characterizes the linearity of laser frequency modulation. The waveform is shown in the third row of Fig.~\ref{Extended_Fig3} (also in Fig.~\ref{Fig3}a and b of the main text). As can be seen from the figure, the net deviation is less than 10 \% at all measured frequencies, indicating a high linearity of laser frequency modulation over a large frequency tuning range. Moreover, of one's preference, a higher linearity can be achieved by sacrificing the tuning range as shown in the first column of extended Fig.~\ref{Extended_Fig3}, where a 3 \% deviation is obtained with a 1.2-GHz detuning range. 

\medskip

\begin{figure}[t!]
	\includegraphics[width=1.0\columnwidth]{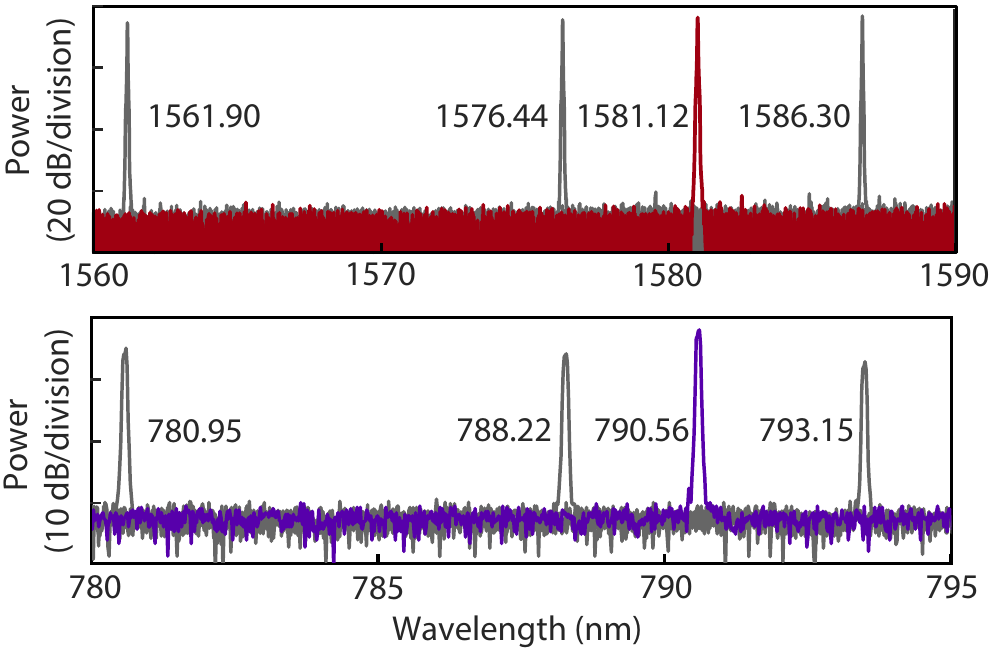}
	\caption{\label{Extended_Fig4} {\bf Optical spectra of the fundamental-frequency lasing mode (top) and the up-converted light at the second harmonic (bottom).} The measured signal at different wavelength of fundamental-frequency lasing mode (top) and generated second-harmonic mode (botton) from the same device, indicating a dual lasing bandwidth over ~10 nm at visible wavelength.}
\end{figure}

\noindent\rmfamily\textbf{Switching performance characterization.} The lasing mode switching behavior is explored by the setup shown in Fig.~\ref{Extended_Fig2}. A square wave is produced by a function generator (Keysight 22621A) and applied to the electrodes sitting beside the resonator. By controlling the wavelength mismatch between two resonators, the lasing mode can be switched from one to another at minimum and maximum values of the square wave respectively as shown in Fig.~\ref{Fig3}g. The lasing signal is then divided by a WDM and converted to electrical signal by two independent photodetector and recorded by a real-time oscilloscope. 

The dual-wavelength switching experiment is implemented with the similar setup via replacing the WDM and one of the photodetector by a visible band one (Thorlabs APD130A). For the recorded fundamental switching signals, the rise and fall time are defined by the bandwidth of function generator. In comparison, the ones of SHG light are limited by the bandwidth of the visible band photodetector, which are around ~15 ns. Moreover, in Fig.~\ref{Fig4}e, the observed spikes are signal overshoot/undershoot caused by the same limitation, which are irrelevant to the performance of device.

\medskip

\noindent\rmfamily\textbf{Wide bandwidth dual wavelength laser.} Our laser exhibits wide tunability realized by thermal optical effect, allowing uniform lasing power as show in  Fig.~\ref{Fig2}c and top of Fig.~\ref{Extended_Fig4}. To further achieve wide tuning range at visible wavelength, we tailored the design of PPLN resonator, and obtain a theoretical phase matching window at 1575 nm with over 20 nm bandwidth. By combining these two together, the fundamental and second-harmonic modes in the resonator are quasi-phase matched over a broad band, leading to an extremely strong nonlinearity effect within the cavity. The generated visible lights are observed and recorded from the same device in the bottom of Fig.~\ref{Extended_Fig4}, which proves the wide tunability of our laser at visible band over ~10 nm. The slight power variation of the SH light can be further optimized in the future by modifying the coupling strength at SH resonances of the PPLN resonator.

    \medskip
	
	\noindent\sffamily\textbf{Acknowledgements}
	
	\noindent\rmfamily The authors thank Prof. Hui Wu for the use of his equipment. They also thank Wenhui Hou, Wuxiucheng Wang, Lejie Lu, and Ming Gong for valuable discussions and help on experiment, and Prof. David Weld and Prof. Manuel Endres for discussions on atomic physics. This work is supported in part by the Defense Advanced Research Projects Agency (DARPA) LUMOS program under Agreement No.~HR001-20-2-0044, the Defense Threat Reduction Agency-Joint Science and Technology Office for Chemical and Biological Defense (grant No.~HDTRA11810047), and the National Science Foundation (NSF) (ECCS-1810169, ECCS-1842691 and, OMA-2138174). This work was performed in part at the Cornell NanoScale Facility, a member of the National Nanotechnology Coordinated Infrastructure (National Science Foundation, ECCS-1542081); and at the Cornell Center for Materials Research (National Science Foundation, Grant No. DMR-1719875). \newline

	\noindent\textbf{Author Contributions} M.L., L.C. and Q.L. conceived the experiment. M.L., J.S., U.J. and T.M. performed numerical simulations. M.L., J.L., Y.H. and S.X. fabricated the device. M.L., J.L, R.L and S.Z. carried out the device characterization. L.W., B.S., and H.W. helped on the characterization of laser linewidth. L.Z. provided valuable suggestions to the design of device. M.L. and L.C. wrote the manuscript with contributions from all authors. Q.L., J.B. and K.V. supervised the project. Q.L. conceived the device concept. \newline


	
	\noindent\textbf{Competing interests.} The authors declare no competing interests.  \newline
	
	\noindent\textbf{Data and materials availability} All data are available in the main text or the supplementary materials. Correspondence and requests for materials should be sent to Q.L. (qiang.lin@rochester.edu), J.B. (jbowers@ucsb.edu ), and K.V. (vahala@caltech.edu).
	
\end{document}


\title{\bf Supplementary}

\author{Mingxiao Li}
\thanks{These two authors contributed equally}
\affiliation{Department of Electrical and Computer Engineering, University of Rochester, Rochester, NY 14627}
\author{Lin Chang}
\thanks{These two authors contributed equally}
\affiliation{Department of Electrical and Computer Engineering, University of California Santa Barbara, Santa Barbara, CA 93106}
\author{Lue Wu}
\affiliation{T. J. Watson Laboratory of Applied Physics, California Institute of Technology, Pasadena, CA 91125}
\author{Jeremy Staffa}
\affiliation{Institute of Optics, University of Rochester, Rochester, NY 14627}	
\author{Jingwei Ling}
\affiliation{Department of Electrical and Computer Engineering, University of Rochester, Rochester, NY 14627}
\author{Usman A. Javid}
\affiliation{Institute of Optics, University of Rochester, Rochester, NY 14627}
\author{Yang He}
\affiliation{Department of Electrical and Computer Engineering, University of Rochester, Rochester, NY 14627}
\author{Raymond Lopez-rios}
\affiliation{Department of Electrical and Computer Engineering, University of Rochester, Rochester, NY 14627}
\author{Shixin Xue}
\affiliation{Department of Electrical and Computer Engineering, University of Rochester, Rochester, NY 14627}	
\author{Theodore J. Morin}
\affiliation{Department of Electrical and Computer Engineering, University of California Santa Barbara, Santa Barbara, CA 93106}
\author{Boqiang Shen}
\affiliation{T. J. Watson Laboratory of Applied Physics, California Institute of Technology, Pasadena, CA 91125}
\author{Heming Wang}
\affiliation{T. J. Watson Laboratory of Applied Physics, California Institute of Technology, Pasadena, CA 91125}
\author{Siwei Zeng}
\affiliation{Department of Electrical and Computer Engineering, Center for Optical Materials Science and Engineering Technologies, Clemson University, Clemson, SC 29634}
\author{Lin Zhu}
\affiliation{Department of Electrical and Computer Engineering, Center for Optical Materials Science and Engineering Technologies, Clemson University, Clemson, SC 29634}
\author{Kerry J. Vahala}

\affiliation{T. J. Watson Laboratory of Applied Physics, California Institute of Technology, Pasadena, CA 91125}
\author{John E. Bowers}

\affiliation{Department of Electrical and Computer Engineering, University of California Santa Barbara, Santa Barbara, CA 93106}
\author{Qiang Lin}

\affiliation{Department of Electrical and Computer Engineering, University of Rochester, Rochester, NY 14627}
\affiliation{Institute of Optics, University of Rochester, Rochester, NY 14627}



\maketitle

\noindent\sffamily\textbf{Introduction}\nolinebreak
	
	\noindent\rmfamily Modern signal processing needs fully-integrated lasers within an optical communication system. Currently,  integrated narrow-linewidth lasers lack fast tuning as well as frequency conversion capability, which limits many applications in integrated photonics \cite{Nokia20}. Lithium niobate (LiNbO$_3$, LN), as a well known material with strong second-order nonlinearity, has proved its capability in high quality modulation in optical communication systems with high speed signal control. Moreover, the rapid progress and many explorations in thin film LN have extended the range of feasible applications, especially in the area of wavelength conversion. And this stimulates interest in full integration of LN functions with other required functions such as optical gain \cite{Loncar18_2,Usman21}.\newline\linebreak 

\noindent\sffamily\textbf{Results}\nolinebreak
	
\noindent\rmfamily\textbf{Device design.}

	Between the RSOA and LN chip, the mode mismatch causes enormous insertion loss, which degrades the output power of laser. As a result, we introduce a spot-size converter to the system, and obtain a minimal mode mismatch when implement an input waveguide with 5 ${\mu}$m width and 600 nm thickness of LN thin film, simulated by FEM software, details showed in Method \cite{Bergman18}. The input waveguide on LN chip is also angled to minimize the coupling loss.

	The lasing linewidth is filtered and narrowed by Vernier effect formed via a dual racetrack structure. The width and thickness of racetracks and bus waveguides are tailored to minimize the number of coupled mode families, here, only the fundamental TE mode, to avoid multi-mode lasing in the structure. A slightly over-coupled regime is preferred to enhance the lasing power while limiting the linewidth broadening caused by the loading quality factor degradation. After optimization, a uniform and broad band coupling is obtained, realizing a consistent lasing power at different wavelengths. Moreover, the racetrack structure is selected to substitute the widely used ring resonator. Since the polarization of mode can be aligned to maximize the effective mode volume utilizing the $r_{33}$ tensor in LN, accordingly, leading to a higher effective EO factor in the structure. Additionally, the value of racetrack radius considers the trade-off between optical scattering loss and EO tuning efficiency. The perimeter difference between two racetracks are also selected for high extinction ratio, reasonable free spectral range (FSR) and Vernier FSR, which is 70 GHz and 2.4 THz respectively in this case \cite{BowersTutorial19}. 
	
	On the other hand, each of the racetracks is separately controlled by different functionalities. The first racetrack is incorporated with a micro heater, realizing the coarse wavelength tuning. While the second one is EO operated, introducing a high speed (sub-GHz level) fine tuning capability to the system. After that, a Sagnac loop ring is placed to control the reflection of output mirror, meanwhile, forming the lasing cavity. A lensed fiber is used to couple the light from tapered fiber for performance characterization, as shown in Fig~\ref{Fig1(b)}.
	
    Additionally, the outstanding material properties of LN in nonlinear optics enable the path to multicolor laser by utilization of periodically poled lithium niobate (PPLN) thin film. Here, we tailor the design to be compatible with harmonic generation applications and pole one of the racetrack with short pulses. The applied period enables a phase matching window from 1560 nm to 1600 nm, which allows us to explore the switching nautre of the wavelength converter. The design, mathematical modeling and preparation of the PPLN racetrack are discussed in Supplementary.
	
\medskip
	
\noindent\textbf{Linear performance.} 
	
	Fig.~\ref{Fig1}(c) shows a fabricated device (see Methods for the details of device fabrication). The design of the metallic structure considers the impedance matching and group velocity match between electrical and optical signals, and optimized to minimize the coupling loss of the RF signal from the probes to the device. 
	
	The laser performance is tested with the mode at 1581.12 nm, which has a threshold current of 80 mA and a maximum on-chip power of around 3.7 mW at 200 mA, accounting for the 5 dB coupling loss from the facet. The highest on-chip power measured from this device is more than 5.5 mW at another Vernier lasing mode.

	The phase noise of the laser is measured by correlated delayed self-heterodyne phase noise measurement \cite{Vahala21,Linewidth08}.
	The experiment setup is shown in Fig.~\ref{Fig2}. The lasing signal passed a 50:50 splitter. One of the optical path is delayed by an 1 km fiber line, while another path of signal is modulated by an acoustic optic modulator (AOM) at 55 MHz. After that, the signals are combined by another splitter and detected by a photodetector. Finally, the signal is analyzed by a real-time oscilloscope.
	
    The measured phase noise is shown in Fig.~\ref{Fig2}. The frequency noise is calculated from the phase noise, as shown in the inset of Fig.~\ref{Fig2}. The white noise floor is found at noise frequency around 10 MHz, which is about 2 ${KHz^2/Hz}$. The extracted phase noise at higher frequencies is relatively free from technical noise from electronics, vibrations, and other environmental factors. This value can be multiplied by 2${\pi}$ to indicate the intrinsic linewidth, or Lorentzian linewidth of the laser, which is 12.6 KHz in this case. 

	\begin{figure*}[t!]
		\centering\includegraphics[width=1.0\columnwidth]{Figure_2.pdf}
		\caption{\label{Fig2} {\bf (a)} Schematic of the hybrid integrated laser structure with LIV curve. {\bf (b)} Setup schematic of delayed self-heterodyne phase noise measurement. The length of delay line is selected for different experiments. AOM: acoustic optical modulator. OSC: real-time oscilloscope. {\bf (c)} Noise spectra of laser measured by non-correlated self-heterodyne measurement. The curve is fitted by a combination of Lorentzian and Gaussian distribution. {\bf (d)} Noise spectra of laser measured by correlated delayed self-heterodyne measurement. The dark blue line is the phase noise signal measured by a real-time oscilloscope. The light blue line shows the frequency noise signal derived from the phase noise. A white noise floor is highlighted by the dashed red line correspondingly in both plots. The dashed rectangular indicates the frequency range used in frequency noise signal calculation. {\bf (e)} Thermal-optic tuning of the laser wavelength measured by an optical spectrum analyzer. The red arrow highlights the lasing mode at 1581.12 nm used to characterize the performance of this device.}
	\end{figure*}
	
\medskip
	
\noindent\textbf{Nonlinear tunability.}

	Tunability of a laser paves the road for numerous applications in nanophotonics. Here we implement both coarse and fine tunability in our laser, while the coarse one is realized by thermo-optic (TO) effect as shown in Fig.~\ref{Fig2}(e). With the integrated micro heater on the racetrack, we are able to apply a direct current signal to the heater, resulting in the increase of temperature and causing the blue shift of lasing wavelength in the spectrum via the thermo-optic effect. The measured free spectral range (FSR) is roughly 70 GHz, while the Vernier FSR is 2.4 THz corresponding to 20 nm from 1576 nm to 1596 nm \cite{BowersTutorial19}.

	\begin{figure*}[t!]
		\includegraphics[width=1.0\columnwidth]{Figure_3.pdf}
		\caption{\label{Fig3}  {\bf (a)} Measured FFT signal from the beat note between the lasing signal and an external cavity diode laser (ECDL) with a modulating signal applied to the phase shifter at V$_p$ = 0 V, 0.6V, 1.8 V, 3.0V and 5.0 V, respectively. The insets in (a,b) show the schematic of the beatnote with and without modulation. {\bf (b)} Processed STFT signal from the beat note moudulated by a triangular function at frequency = 10 MHz, 50 MHz, 100 MHz, 200MHz respectively and V$_p$ = 3 V. {\bf (c)} The signal recorded from the real-time oscilloscope. The upper one shows the signal with a duration of 1 ${\mu}$s at modulation speed of 50 MHz. The lower one is the zoomed in figure of one modulation period. {\bf (d)} Recorded switching signal from two adjacent mode in the laser at 10 MHz and 50 MHz respectively.}
	\end{figure*}

	Additionally, due to the large electro-optic (EO) factor in LN, we are able to realize fine frequency tuning of the wavelength via the EO effect. As shown in Fig.~\ref{Fig2}(a), we use an external cavity diode laser (ECDL) to beat with the lasing signal at a frequency difference of 3.1 GHz. Then we modulate the lasing cavity with a RF signal applied to the phase shifter at 50 MHz. The beat note is modulated as shown in inset of the second panel in Fig.~\ref{Fig2}(a), and the tuning range derived from the fast fourier transform (FFT) is shown in Fig.~\ref{Fig2}(a) at V$_p$=0.6, 1.8 V, 3.0 V, and 5.0 V, respectively. The maximum tuning efficiency is found at V$_p$=3.0 V, indicating a rate of 330 MHz/V at radio frequency, while a maximum mode hoping free tuning range of 2.48 GHz is obtained at 50 MHz with a lower efficiency, since it approaches the tuning limit.
	
	For a better indication of the fine tuning performance via the Pockels effect, we implement the triangular modulation to the device. The signal generated from a arbitrary waveform generator (AWG) ranging from 10 MHz to 200 MHz at V$_p$=3.0 V. The heterodyne beat note is recorded and processed by short time fourier transform (STFT). As shown in Fig.~\ref{Fig2}(b), the frequency detuning is around 1.98 GHz, which corresponds to the derived tuning efficiency previously. The efficiency degrades a little bit at higher frequency due to extra loss from electrodes. This result indicates the non-degradation modulation of the linearity of the input signal up to 200 MHz. The resolution of the figure is limited by the bandwidth and sampling rate of the used realtime oscilloscope. 
	
	With an efficient and high-quality tuning capability, we are able to realize a switching of lasing mode at tens of MHz. We apply a square signal at 10 MHz and 50 MHz. As shown in Fig.~\ref{Fig2}(a), the switching of laser mode is obtained at respectively. The speed is limited by the oscillation nature in the laser during cavity mode stabilization \cite{Nokia20}. This switching capability not only realizes the fast wide modulation of the lasing wavelength, but also enables the application for light switch in a wavelength converter, to be discussed in the next section.
	
	\medskip
	
	\noindent\textbf{Wavelength converter switch.} 
	
	Conventional integrated lasers have incorporated second harmonic generation (SHG) outside of the laser cavity, while here we place periodically-poled lithium niobate (PPLN) into the laser cavity to achieve more efficient SHG, due to the higher intracavity power (40\% higher in this device). The design, mathematical modeling and preparation of the PPLN racetrack are discussed in Supplementary.

	\begin{figure*}[t!]
		\includegraphics[width=1.0\columnwidth]{Figure_4.pdf}
		\caption{\label{Fig4}  {\bf (a)} The measured FH and SHG signal from OSA.{\bf (b)} The generated SHG light emission observed under a microscope with visible camera at the facet. {\bf (c)} Measured frequency doubling signal generated by quasi-phase matching at various lasing power as shown in Fig.~\ref{Fig2}(a), with quadratical fitting. }
	\end{figure*}
	
	After the poling of the device, with the device lasing at 1581.12 nm, a SHG signal is detected at 
	790.56 nm, which is half of the lasing wavelength, as shown in Fig.~\ref{Fig4}(a). The emitted SHG signal is also observed under a visible light camera, as shown in Fig.~\ref{Fig4}(b). 
	Moreover, we utilize both the fundamental lasing power and SHG output light data recorded at LIV curve measurement and verifies the quadratic dependence as fitted in Fig.~\ref{Fig4}(c), which indicates the intrinsic nature of second order nonlinearity. 
	
	Now the wavelength converter is embedded in the laser with EO tuning capability. To exhibit the switch performance, first, we match the fundamental harmonic (FH) and second harmonic (SH) resonance by tuning the temperature via the thermal electrical controller (TEC) on the setup, maximizing the conversion efficiency. Then, similar to our previosu test process, we apply a square signal to modulate the lasing cavity, which introduce a frequency mismatch between FH and SH resonance, leading to a on-off mechanism. As shown in Fig.~\ref{Fig4}(d), the switch nature is observed at 10 MHz and 50 MHz respectively. \newline\linebreak

	\noindent\sffamily\textbf{Discussion}\nolinebreak
	
	\noindent\rmfamily In summary, by hybrid integration of a LN external cavity with a III-V RSOA, we demonstrated the first integrated nonlinear laser, which exhibits both fast EO tuning with efficiency of 330 MHz/V and modulating speed more than 200 MHz, as well as intracavity frequency conversion with microwatts output. By combining these two functionalities, an application in fast switching wavelength converter is demonstrated up to 50 MHz.
	
	Additionally, the outstanding nonlinear-optical properties of LN can extend these results to enable multicolor fully chip integrated lasers. More generally, the demonstration of this laser paves the way to fully integrated circuits incorporating gain and LN functionality. These circuits can be used in optical signal processing systems as well as various applications in nonlinear optics, quantum photonics and optical communications.
	
    \bigskip
    
    \noindent\sffamily\textbf{Methods}\nolinebreak
    
    \noindent\rmfamily\textbf{Device fabrication.} The devices were fabricated on a $600$-nm-thick x-cut single-crystalline LN thin film bonded on a 4.7-${\mu}$m silicon dioxide layer sitting on a silicon substrate (from NanoLN). The waveguide and racetrack structures are patterned by ZEP-$520$A positive resist via the electron-beam lithography, an Ar$^+$ plasma milling process is used to  transfer the pattern to the LN layer with etch depth of 300-nm. The resist was removed by the solvent 1165 resist remover afterwards. The metal electrode layer (10 nm Ti/500 nm Au) was patterned by PMMA and deposited by an electron-beam evaporator. Then formed by an overnight lift-off process. The device is diced and the facets are polished to minimize the edge coupling loss.